\pdfoutput=1

\documentclass[11pt]{article}

\usepackage{naacl2021}

\usepackage{times}
\usepackage{latexsym}
\usepackage{graphicx}
\usepackage{booktabs}
\usepackage{subfig}
\usepackage[ruled,vlined]{algorithm2e}

\usepackage[T1]{fontenc}

\usepackage[utf8]{inputenc}

\usepackage{microtype}

\title{\textit{Query2Prod2Vec} \\ Grounded Word Embeddings for eCommerce}

\author{Federico Bianchi\\
  Bocconi University\\
  Milano, Italy \\
  f.bianchi@unibocconi.it \\\And
  Jacopo Tagliabue\thanks{\textbf{ }Corresponding author. All authors contributed equally and are listed alphabetically.}\\
  Coveo Labs\\
  New York, USA\\
  jtagliabue@coveo.com \\\And
  Bingqing Yu \\
  Coveo\\
  Montreal, Canada\\
  cyu2@coveo.com \\
  }

\begin{document}
\maketitle
\begin{abstract}
   We present~\textbf{Query2Prod2Vec}, a model that grounds lexical representations for product search in product embeddings: in our model,~\textit{meaning} is a mapping between words and a latent space of products in a digital shop. We leverage shopping sessions to learn the underlying space and use merchandising annotations to build lexical analogies for evaluation: our experiments show that our model is more accurate than known techniques from the NLP and IR literature. Finally, we stress the importance of data efficiency for product search outside of retail giants, and highlight how~\textbf{Query2Prod2Vec} fits with practical constraints faced by most practitioners.
\end{abstract}

\section{Introduction}
\label{sec:intro}

The eCommerce market reached in recent years an unprecedented scale: in 2020, 3.9 trillion dollars were spent globally in online retail~\cite{emarketer2020}. While shoppers make significant use of search functionalities, improving their experience is a never-ending quest~\cite{econsultancy2020}, as outside of few retail giants users complain about sub-optimal performances~\cite{Baymard2020}. As the technology behind the industry increases in sophistication, neural architectures are gradually becoming more common~\cite{Tsagkias2020ChallengesAR} and, with them, the need for accurate word embeddings for Information Retrieval (IR) and downstream Natural Language Processing (NLP) tasks~\cite{Yu2020BlendingSA,CoveoECNLP22}. 

Unfortunately, the success of standard and contextual embeddings from the NLP literature~\cite{Mikolov2013EfficientEO,bert} could not be immediately translated to the product search scenario, due to some peculiar challenges~\cite{bianchi2020bert}, such as short text, industry-specific jargon~\cite{Bai:2018:SQN:3219819.3219897}, low-resource languages; moreover, specific embedding strategies have often been developed in the context of high-traffic websites~\cite{Grbovic2016ScalableSM}, which limit their applicability in many practical scenarios. In \textit{this} work, we propose a sample efficient word embedding method for IR in eCommerce, and benchmark it against SOTA models over industry data provided by partnering shops. We summarize our contributions as follows:

\begin{enumerate}
    \item we propose a method to learn dense representations of words for eCommerce: we name our method~\textbf{Query2Prod2Vec}, as the mapping between words and the latent space is mediated by the product domain; 
    \item we evaluate the lexical representations learned by~\textbf{Query2Prod2Vec} on an analogy task against SOTA models in NLP and IR; benchmarks are run on two independent shops, differing in traffic, industry and catalog size;
    \item  we detail a procedure to generate synthetic embeddings, which allow us to tackle the ``cold start'' challenge;
    \item  we release our implementations, to help the community with the replication of our findings on other shops\footnote{Public repository available at:~\url{https://github.com/coveooss/ecommerce-query-embeddings}.}.
\end{enumerate}

While perhaps not fundamental to its industry significance, it is important to remark that grounded lexical learning is well aligned with theoretical considerations on~\textit{meaning} in recent (and less recent) literature~\cite{bender-koller-2020-climbing,Bisk2020ExperienceGL,montague1974c}.

\section{Embeddings for Product Search: an Industry Perspective}
\label{sec:industry}

In product search, when the shopper issues a query (e.g. ``sneakers'') on a shop, the shop search engine returns a list of $K$ products matching the query intent and possibly some contextual factor -- the shopper at that point may either leave the website, or click on $n$ products to further explore the offering and eventually make a purchase.

Unlike web search, which is exclusively performed at massive scale, product search is a problem that both big and small retailers have to solve: while word embeddings have revolutionized many areas of NLP~\cite{Mikolov2013EfficientEO}, word embeddings for product queries are especially challenging to obtain at scale, when considering the huge variety of use cases in the overall eCommerce industry. In particular, based on industry data and first-hand experience with dozens of shops in our network, we identify four~\textit{constraints} for effective word embeddings in eCommerce:

\begin{enumerate}
    \item \textbf{Short text}. Most product queries are very short -- 60\% of all queries in our dataset are one-word queries, > 80\% are two words or less; the advantage of contextualized embeddings may therefore be limited, while lexical vectors are fundamental for downstream NLP tasks~\cite{Yu2020BlendingSA,BianchiSIGIReCom2020}. For this reason, the current work specifically addresses the quality of~\textit{word} embeddings\footnote{Irrespectively of how the lexical vectors are computed, query embeddings can be easily recovered with the usual techniques (e.g. sum or average word embeddings~\cite{Yu2020AnII}): as we mention in the concluding remarks, investigating compositionality is an important part of our overall research agenda.}.
    \item \textbf{Low-resource languages}. Even shops that have the majority of their traffic on English domain typically have smaller shops in low-resource languages.
    \item \textbf{Data sparsity}. In~\textit{Shop X} below, only 9\% of all shopping sessions have a search interaction\footnote{This is a common trait verified across industries and sizes: among dozens of shops in our network, 30\% is the highest \textit{search vs no-search} session ratio;~\textit{Shop Y} below is around 29\%.}. Search sparsity, coupled with vertical-specific jargon and the usual long tail of search queries, makes data-hungry models unlikely to succeed for most shops.
    \item \textbf{Computational capacity}. The majority of the market has the necessity to strike a good trade-off between quality of lexical representations and the cost of training and deploying models, both as hardware expenses and as additional maintenance/training costs. 
\end{enumerate}

The embedding strategy we propose --~\textbf{Query2Prod2Vec} -- has been designed to allow efficient learning of word embeddings for product queries. Our findings are useful to a wide range of practitioners: large shops launching in new languages/countries, mid-and-small shops transitioning to dense IR architectures and the raising wave of multi-tenant players\footnote{As an indication of the market opportunity, only in 2019 and only in the space of AI-powered search and recommendations, we witnessed Coveo~\cite{CoveoRound}, Algolia~\cite{AlgoliaRound} and
Lucidworks~\cite{LWRound} raising more than 100M USD each from venture funds.}: as A.I. providers grow by deploying their solutions on multiple shops, ``cold start'' scenarios are an important challenge to the viability of their business model.

\section{Related Work}
\label{sec:related}
The literature on learning representations for lexical items in NLP is vast and growing fast; as an overview of classical methods,~\citet{baroni-etal-2014-dont} benchmarks several count-based and neural techniques~\cite{Landauer1997AST,w2vec}; recently, context-aware embeddings~\cite{elmo,bert} have demonstrated state-of-the-art performances in several semantic tasks~\cite{rogers2020bertology,nozza2020mask}, including document-based search~\cite{Nogueira2020DocumentRW}, in which target entities are long documents, instead of product~\cite{Craswell2020OverviewOT}. 
To address IR-specific challenges, other embedding strategies have been proposed:~\textit{Search2Vec}~\cite{Grbovic2016ScalableSM} uses interactions with ads and pages as context in the typical context-target setting of skip-gram models~\cite{w2vec};~\textit{QueryNGram2Vec} \cite{Bai:2018:SQN:3219819.3219897} additionally learns embeddings for n-grams of word appearing in queries to better cover the long tail. The idea of using vectors (from images) as an aid to query representation has also been suggested as a heuristic device by~\citet{Yu2020AnII}, in the context of personalized language models;~\textit{this} work is the first to our knowledge to benchmark embeddings on lexical semantics (not tuned for domain-specific tasks),~\textit{and} investigate sample efficiency for small-data contexts.

\section{Query2Prod2Vec}
\label{sec:embeddings}

In~\textbf{Query2Prod2Vec}, the representation for a query~\textit{q} is built through the representation of the objects that~\textit{q} refers to. Consider a typical shopper-engine interaction in the context of product search: the shopper issues a query, e.g. ``shoes'',  the engine replies with a noisy set of potential referents, e.g. pairs of shoes from the shop inventory, among which the shopper may select relevant items. Hence, this dynamics is reminiscent of a cooperative language game~\cite{lewis1969convention}, in which shoppers give noisy feedback to the search engine on the meaning of the queries. A full specification of~\textbf{Query2Prod2Vec} therefore involves a representation of the target domain of reference (i.e. products in a digital shop) and a denotation function. 

\subsection{Building a Target Domain}
\label{sec:prod2vec}
We represent products in a target shop through a~\textit{prod2vec} model built with anonymized shopping sessions containing user-product interactions. Embeddings are trained by solving the same optimization problem as in classical~\textit{word2vec}~\cite{Mikolov2013EfficientEO}:~\textit{word2vec} becomes~\textit{prod2vec} by substituting \textit{words} in a \textit{sentence} with \textit{products} viewed in a \textit{shopping session}~\cite{DBLP:journals/corr/abs-1804-00306}.
The utility of~\textit{prod2vec} is independently justified~\cite{Grbovic15,Tagliabue2020ShoppingIT} and, more importantly, the referential approach leverages the abundance of browsing-based interactions, as compared to search-based interactions: by learning \textit{product} embeddings from abundant behavioral data first, we sidestep a major obstacle to reliable word representation in eCommerce. Hyperparameter optimization follows the guidelines in~\citet{BianchiSIGIReCom2020}, with a total of $26{,}057$ (\textit{Shop X}) and $84{,}575$ (\textit{Shop Y}) product embeddings available for downstream processing\footnote{Final parameters for~\textit{prod2vec} are: $dimension=50$, $win\_size=10$, $iterations=30$, $ns\_exponent=0.75$.}. 
 
\subsection{Learning Embeddings}
The fundamental intuition of~\textbf{Query2Prod2Vec} is treating clicks after~\textit{q} as a noisy feedback mapping~\textit{q} to a portion of the latent product space. In particular, we compute the embedding for~\textit{q} by averaging the product embeddings of all products clicked after it, using frequency as a weighting factor (i.e. products clicked often contribute more). The model has one free parameter,~\textit{rank}, which controls how many embeddings are used to build the representation for~\textit{q}: if~\textit{rank=k}, only the~\textit{k} most clicked products after~\textit{q} are used. The results in Table~\ref{tab:at_st} are obtained with~\textit{rank=5}, as we leave to future work to investigate the role of this parameter.

The lack of large-scale search logs in the case of new deployments is a severe issue for successful training. The referential nature of~\textbf{Query2Prod2Vec} provides a fundamental competitive advantage over models building embeddings from past linguistic behavior~\textit{only}, as synthetic embeddings can be  generated as long as cheap session data is available to obtain an initial~\textit{prod2vec} model. As detailed in the ensuing section, the process happens in two stages, event generation and embeddings creation. 

\subsection{Creating Synthetic Embeddings}
The procedure to create synthetic embeddings is detailed in Algorithm~\ref{alg:synth}: it takes as input a list of words, a pre-defined number of sampling iterations, a popularity distribution over products\footnote{Please note that data on product popularity can be easily obtained through marketing tools, such as Google Analytics.}, and it returns a list of synthetic search events, that is, a mapping between words and lists of products ``clicked''. Simulating the~\textit{search} event can be achieved through the existing search engine, as, from a practical standpoint,~\textit{some} IR system must already be in place given the use case under consideration. To avoid over-relying on the quality of IR and prove the robustness of the method, all the simulations below are not performed with the actual production API, but with a custom-built inverted index over product meta-data, with a simple TF-IDF weighting and Boolean search. 

\begin{algorithm}

\SetAlgoLined
\SetKwFunction{Search}{Search}\SetKwFunction{Sample}{Sample}
\KwData{a list of words $W$, a pre-defined number $N$ of simulations per word, a distribution $D$ over products. }
\KwResult{A dataset of synthetic clicked events: $E$}
 $E\leftarrow ~\textit{empty mapping}$\;
 
 \ForEach{word $w$ in $W$}{
  product\_list~$\leftarrow$~\Search{$w$}\;
  \For{$i=1$ \KwTo $N$}{
    $p$~$\leftarrow$~\Sample(product\_list, $D$)\;
    append the entry ($w$,~$p$) to $E$;
  }
 }
 \Return $E$\
 \caption{Generation of synthetic click events.}
 \label{alg:synth}
\end{algorithm}

For the second stage, we can treat the synthetic click events produced by Algorithm~\ref{alg:synth} as a drop-in replacement for user-generated events -- that is, for any query~\textit{q}, we calculate an embedding by averaging the product embeddings of the relevant products, weighted by frequency\footnote{Please note that while~\textit{this} work focuses on lexical quality, the same strategy can be applied to complex queries in a ``cold start'' scenario.}. Putting the two stages together,~\textbf{Query2Prod2Vec} can not only produce reliable query embeddings based on historical data, but also learn approximate embeddings for a large vocabulary~\textit{before} being exposed to any search interaction: in Section~\ref{sec:results} we report the performance of~\textbf{Query2Prod2Vec} when using only synthetic embeddings\footnote{All the experiments are performed with $N=500$ simulated search events per query.}.

\section{Dataset and Baselines}
\label{sec:exp}

\subsection{Dataset}
\label{sec:dataset}
Following best practices in the multi-tenant literature~\cite{10.1145/3383313.3411477}, we benchmark all models on different shops to test their robustness. In particular, we obtained catalog data, search logs and anonymized shopping sessions from two partnering shops,~\textit{Shop X} and~\textit{Shop Y}:~\textit{Shop X} is a sport apparel shop with Alexa ranking of approximately~\textit{200k}, representing a prototypical shop in the middle of the long tail;~\textit{Shop Y} is a home improvement shop with Alexa ranking of approximately~\textit{10k}, representing an intermediate size between~\textit{Shop X} and public companies in the space. Linguistic data is in Italian for both shops, and training is done on random sessions from the period June-October 2019: after sampling, removal of bot-like sessions and pre-processing, we are left with $722{,}479$ sessions for~\textit{Shop X}, and $1{,}986{,}452$ sessions for~\textit{Shop Y}.

\subsection{Baselines}
\label{sec:base}
We leverage the unique opportunity to join catalog data, search logs and shopping sessions to extensively benchmark~\textbf{Query2Prod2Vec} against a variety of methods from NLP and IR. 

\begin{itemize}
    \item \textbf{Word2Vec and FastText}.~We train a CBOW~\cite{Mikolov2013EfficientEO} and a FastText model~\cite{bojanowski2016enriching} over product descriptions in the catalog;
    \item \textbf{UmBERTo}.~We use RoBERTa trained on Italian data  -- \textit{UmBERTo}\footnote{\url{https://huggingface.co/Musixmatch/umberto-commoncrawl-cased-v1}}. The $\langle s \rangle$ embedding of the last layer of the architecture is the query embedding;
    \item \textbf{Search2Vec}.~We implement the skip-gram model from~\citet{Grbovic2016ScalableSM}, by feeding the model with sessions composed of search queries and user clicks. Following the original model, we also train a time-sensitive variant, in which time between actions is used to weight query-click pairs differently;
    \item \textbf{Query2Vec}.~We implement a different context-target model, inspired by~\citet{Grubhub}: embeddings are learned by the model when it tries to predict a (purchased or clicked) item starting from a query;
    \item \textbf{QueryNGram2Vec}.~We implement the model from~\citet{Bai:2018:SQN:3219819.3219897}. Besides learning representations through a skip-gram model as in~\citet{Grbovic2016ScalableSM}, the model learns the embeddings of \textit{unigrams} to help cover the long tail for which no direct embedding is available.
\end{itemize}

To guarantee a fair comparison, all models are trained on the same sessions. For all baselines, we follow the same hyperparameters found in the cited works: the dimension of query embedding vectors is set to 50, except that 768-dimensional vectors are used for~\textbf{UmBERTo}, as provided by the pre-trained model. 

As discussed in Section~\ref{sec:intro}, a distinguishing feature of~\textbf{Query2Prod2Vec} is~\textit{grounding}, that is, the relation between words and an external domain -- in this case, products. It is therefore interesting not only to assess a possible \textit{quantitative} gap in the quality of the representations produced by the baseline models, but also to remark the~\textit{qualitative} difference at the core of the proposed method: if words are~\textit{about} something, pure co-occurrence patterns may be capturing only fragments of lexical meaning~\cite{coveoNAACL21}.

\section{Solving Analogies in eCommerce}
\label{sec:semantic}
As discussed in Section~\ref{sec:industry}, we consider evaluation tasks focused on~\textit{word meaning}, without using product-based similarity (as that would implicitly and unfairly favor referential embeddings).
Analogy-based tasks~\cite{Mikolov2013EfficientEO} are a popular choice to measure semantic accuracy of embeddings, where a model is asked to fill templates like~\textit{man : king = woman : ?}; however, preparing analogies for digital shops presents non trivial challenges for human annotators: these would in fact need to know both the language and the underlying space (``air max'' is closer to ``nike'' than to ``adidas''), with the additional complication that many candidates may not have ``determinate'' answers (e.g. if~\textit{Adidas} is to~\textit{Gazelle}, then~\textit{Nike} is to what exactly?). In building our testing framework, we keep the intuition that analogies are an effective way to test for lexical meaning and the assumption that human-level concepts should be our ground truth: in particular, we programmatically produce analogies by leveraging existing human labelling, as indirectly provided by the merchandisers who built product catalogs\footnote{It is important to note that this categorization is done by product experts for navigation and inventory purposes: all product labels are produced independently from any NLP consideration.}. 

\subsection{Test Set Preparation}
\label{supp:analogy}
We extract words from the merchandising taxonomy of the target shops, focusing on three most frequent fields in query logs:~\textit{product type},~\textit{brand} and~\textit{sport activity} for~\textit{Shop X};~\textit{product type},~\textit{brand} and~\textit{part of the house} for~\textit{Shop Y}. Our goal is to go from taxonomy to analogies, that is, showing how for each pair of taxonomy~\textbf{types} (e.g.~\textit{brand : sport}), we can produce two pairs of~\textbf{tokens} (\textit{Wilson : tennis},~\textit{Cressi : scubadiving}), and create two analogies: \textit{b1 : s1 = b2 : ? (target: s2)} and \textit{b2: s2 = b1 : ? (target: s1)} for testing purposes. For each~\textbf{type} in a pair (e.g.~\textit{brand : sport}), we repeat the following for all possible values of \textit{brand} (e.g.\ ``Wilson'', ``Nike'') -- given a brand $B$:

\begin{enumerate}
    \item we loop over the catalog and record all values of~\textit{sport}, along with their frequency, for the products made by~$B$. For example, for $B=Nike$, the distribution may be: \{``soccer'': 10, ``basketball'': 8, ``scubadiving'': 0 \}; for $B=Wilson$, it may be: \{``tennis'': 8\};
    \item we calculate the Gini coefficient~\cite{gini} over the distribution on the values of~\textit{sport} and choose a conservative Gini threshold, i.e. $75th$ percentile: the goal of this threshold is to avoid ``undetermined'' analogies, such as \textit{Adidas : Gazelle = Nike : ?}. The intuition behind the use of a dispersion measure is that product analogies are harder if the relevant label is found across a variety of products\footnote{In other words,~\textit{Wilson : tennis = Atomic : ? (skiing)} is a better analogy than~\textit{Adidas : Gazelle = Nike : ?}.}.
\end{enumerate}

With all the Gini coefficients and a chosen threshold, we are now ready to generate the analogies, by repeating the following for all values of~\textit{brand} -- given a brand~$B$ we can repeat the following sampling process~$K$ times ($K=10$ for our experiments):

\begin{enumerate}
    \item if~$B$'s Gini value for its distribution of \textit{sport} labels is below our chosen threshold, we skip~$B$; if~$B$'s value is \textit{above}, we associate to~$B$ its most frequent~\textit{sport} value, e.g.~\textit{Wilson : tennis}. This is the \textit{source} pair of the analogy; to generate a \textit{target} pair, we sample randomly a brand~$C$ with high Gini together with its most frequent value, e.g.~\textit{Atomic : skiing};
    \item we add to the final test set two analogies: \textit{Wilson : tennis = Atomic : ?}, and \textit{Atomic : skiing = Wilson : ?}.
\end{enumerate}

\begin{table*}[!ht]
\centering

\begin{tabular}{lcccc}  \toprule
   \textit{Model} & \textbf{HR@5,10 for X} & \textbf{HR@5,10 for Y} & \textbf{CV (X/Y)} & \textbf{Acc on ST}\\ \midrule
   \textit{Query2Prod2Vec (real data)}  & \textbf{0.332} / \textbf{0.468} & \textbf{0.277} / \textbf{0.376} & 0.965/0.924 & \textbf{0.88}\\ \midrule
   \textit{Word2Vec} & 0.206 / 0.242 & 0.005 / 0.009 & 0.47 / 0.03 & 0.68 \\
   \textit{Query2Vec} & 0.077 / 0.113 & 0.065 / 0.120 &  0.97 / 0.93 & 0.54\\
   \textit{QueryNGram2Vec} & 0.071 / 0.122 & 0.148 / 0.216 & \textbf{0.99} / 0.92 & 0.82\\
   \textit{FastText} & 0.068 / 0.116 & 0.010 / 0.012 & 0.52 / 0.03 & 0.57\\
   \textit{UmBERTo} & 0.019 / 0.042 & 0.030 / 0.103  & 0.99 / \textbf{1.00} & 0.57\\
   \textit{Search2Vec (time)} & 0.018 / 0.025 & 0.232 / 0.329 & 0.23 / 0.90  & 0.17\\
   \textit{Search2Vec} & 0.016 / 0.024 & 0.095 / 0.150 & 0.23 / 0.90 &  0.17\\
   \bottomrule

\end{tabular}

\caption{Hit Rate (HR) and coverage (CV) for all models and two shops on~\textbf{AT}; on the rightmost column, Accuracy (Acc) for all models on~\textbf{ST}.}
\label{tab:at_st}
\end{table*}

The procedure is designed to generate test examples conservatively, but of fairly high quality, as for example~\textit{Garmin : watches = Arena : bathing cap} (the analogy relates two brands which sell only one type of items), or~\textit{racket : tennis = bathing cap : indoor swimming} (the analogy relates ``tools'' that are needed in two activities). A total of $1208$ and $606$ test analogies are used for the analogy task~(\textbf{AT}) for, respectively,~\textit{Shop X} and~\textit{Shop Y}: we benchmark all models by reporting Hit Rate at different cutoffs~\cite{Vasile2016MetaProd2VecPE}, and we also report how many analogies are covered by the lexicon learned by the models (\textit{coverage} is the ratio of analogies for which all embeddings are available in the relevant space).

\section{Results}
\label{sec:results}
Table~\ref{tab:at_st} reports model performance for the chosen cutoffs.~\textbf{Query2Prod2Vec} (as trained on real data) has the best performance\footnote{HR@20 was also computed, but omitted for brevity as it confirmed the general trend.}, while maintaining a very competitive coverage. More importantly, following our considerations in Section~\ref{sec:industry}, results confirm that producing competitive embeddings on shops with different constraints is a challenging task for existing techniques, as models tend to either rely on specific query distribution (e.g.~\textbf{Search2Vec (time)}), or the availability of linguistic and catalog resources with good coverage (e.g. \textbf{Word2Vec}).~\textbf{Query2Prod2Vec} is the only model performing with comparable quality in the two scenarios, further strengthening the methodological importance of running benchmarks on more than one shop if findings are to be trusted by a large group of practitioners. 

\subsection{Sample Efficiency and User Studies}
To investigate sample efficiency, we run two further experiments on~\textit{Shop X}: first, we run~\textbf{AT} giving only 1/3 of the original data to~\textbf{Query2Prod2Vec} (both for the \textit{prod2vec} space, and for the denotation). The small-dataset version of~\textbf{Query2Prod2Vec} still outperforms all other full-dataset models in Table~\ref{tab:at_st} (\textit{HR@5,10} = \textit{0.276 / 0.380}). Second, we train a~\textbf{Query2Prod2Vec} model~\textit{only with simulated data} produced as explained in Section~\ref{sec:embeddings} -- that is, with~\textit{zero} data from real search logs. The entirely simulated~\textbf{Query2Prod2Vec} shows performance competitive with the small-dataset version (\textit{HR@5,10} =~\textit{0.259 / 0.363})\footnote{A similar result was obtained on~\textit{Shop Y}, and it is omitted for brevity.}, outperforming all baselines. 

As a further independent check, we supplement~\textbf{AT} with a small semantic similarity task~(\textbf{ST}) on~\textit{Shop X}\footnote{\textit{Shop X} is chosen since it is easier to find speakers familiar with sport apparel than DIY items.}: two native speakers are asked to solve a small set ($46$) of manually curated questions in the form: ``Given the word \textit{Nike}, which is the most similar, \textit{Adidas} or \textit{Wilson}?''.~\textbf{ST} is meant to (partially) capture how much the embedding spaces align with lexical intuitions of generic speakers, independently of the product search dynamics. Table~\ref{tab:at_st} reports results treating human ratings as ground truth and using cosine similarity on the learned embeddings for all models\footnote{Inter-rater agreement was substantial, with~\textit{Cohen Kappa Score}=0.67~\cite{articleKappa}.}.~\textbf{Query2Prod2Vec} outperforms all other methods, further suggesting that the representations learned through referential information capture some aspects of lexical knowledge.

\subsection{Computational Requirements} 
As stressed in Section~\ref{sec:industry}, accuracy and resources form a natural trade-off for industry practitioners. Therefore, it is important to highlight that, our model is not just more accurate, but significantly more efficient to train: the best performing~\textbf{Query2Prod2Vec} takes 30 minutes (CPU only) to be completed for the larger~\textit{Shop Y}, while other competitive models such as~\textbf{Search2Vec(time)} and~\textbf{QueryNGram2Vec} require 2 to 4 hours\footnote{Training is performed on a~\textit{Tesla V100 16GB} GPU. As a back of the envelope calculation, training~\textbf{QueryNGram2Vec} on a~\textit{AWS p3 large} instance costs around 12 USD, while a standard CPU container for~\textbf{Query2Prod2Vec} costs less than 1 USD.}. Being able to quickly generate many models allows for cost-effective analysis and optimization; moreover, infrastructure cost is heavily related to ethical and social issues on energy consumption in NLP~\cite{strubell2019energy}. 

\section{Conclusion and Future Work}

In~\textit{this} work, we learned reference-based word embeddings for product search:~\textbf{Query2Prod2Vec} significantly outperforms other embedding strategies on lexical tasks, and consistently provides good performance in small-data and zero-data scenarios, with the help of synthetic embeddings. In future work, we will extend our analysis to~\textit{i}) specific IR tasks, within the recent paradigm of the dual encoder model~\cite{Karpukhin2020DensePR}, and~\textit{ii}) compositional tasks, trying a systematic replication of the practical success obtained by~\citet{Yu2020AnII} through image-based heuristics. 

When looking at models like~\textbf{Query2Prod2Vec} in the larger industry landscape, we hope our methodology can help the field broaden its horizons: while retail giants indubitably played a major role in moving eCommerce use cases to the center of NLP research, finding solutions that address a larger portion of the market is not just practically important, but also an exciting agenda of its own.

\section{Ethical Considerations}
\textit{Coveo} collects anonymized user data when providing its business services in full compliance with existing legislation (e.g. GDPR). The training dataset used for all models employs anonymous UUIDs to label events and sessions and, as such, it does not contain any information that can be linked to shoppers or physical entities; in particular, data is ingested through a standardized client-side integration, as specified in our public protocol\footnote{\url{https://docs.coveo.com/en/l29e0540/coveo-for-commerce/tracking-commerce-events-reference}}.

\section*{Acknowledgements}

We wish to thank Nadia Labai, Patrick John Chia, Andrea Polonioli, Ciro Greco and three anonymous reviewers for helpful comments on previous versions of this article. The authors wish to thank~\textit{Coveo} for the support and the computational resources used for the project. Federico Bianchi is a member of the Bocconi Institute for Data Science and Analytics (BIDSA) and the Data and Marketing Insights (DMI) unit.

\bibliography{anthology,custom}

\begin{thebibliography}{40}
\expandafter\ifx\csname natexlab\endcsname\relax\def\natexlab#1{#1}\fi

\bibitem[{Bai et~al.(2018)Bai, Ordentlich, Zhang, Feng, Ratnaparkhi, Somvanshi,
  and Tjahjadi}]{Bai:2018:SQN:3219819.3219897}
Xiao Bai, Erik Ordentlich, Yuanyuan Zhang, Andy Feng, Adwait Ratnaparkhi, Reena
  Somvanshi, and Aldi Tjahjadi. 2018.
\newblock \href {https://doi.org/10.1145/3219819.3219897} {Scalable query
  n-gram embedding for improving matching and relevance in sponsored search}.
\newblock In \emph{Proceedings of the 24th ACM SIGKDD International Conference
  on Knowledge Discovery \& Data Mining}, KDD '18, pages 52--61, New York, NY,
  USA. ACM.

\bibitem[{Baroni et~al.(2014)Baroni, Dinu, and
  Kruszewski}]{baroni-etal-2014-dont}
Marco Baroni, Georgiana Dinu, and Germ{\'a}n Kruszewski. 2014.
\newblock \href {https://doi.org/10.3115/v1/P14-1023} {Don{'}t count, predict!
  a systematic comparison of context-counting vs. context-predicting semantic
  vectors}.
\newblock In \emph{Proceedings of the 52nd Annual Meeting of the Association
  for Computational Linguistics (Volume 1: Long Papers)}, pages 238--247,
  Baltimore, Maryland. Association for Computational Linguistics.

\bibitem[{{Baymard Institute}(2020)}]{Baymard2020}
{Baymard Institute}. 2020.
\newblock \href {https://baymard.com/ux-benchmark/chronicle} {\emph{Site Search
  for Ecommerce.}}

\bibitem[{Bender and Koller(2020)}]{bender-koller-2020-climbing}
Emily~M. Bender and Alexander Koller. 2020.
\newblock \href {https://www.aclweb.org/anthology/2020.acl-main.463} {Climbing
  towards {NLU}: {On} meaning, form, and understanding in the age of data}.
\newblock In \emph{Proceedings of the 58th Annual Meeting of the Association
  for Computational Linguistics}, pages 5185--5198, Online. Association for
  Computational Linguistics.

\bibitem[{Bianchi et~al.(2021)Bianchi, Greco, and Tagliabue}]{coveoNAACL21}
Federico Bianchi, Ciro Greco, and Jacopo Tagliabue. 2021.
\newblock {Language in a (Search) Box: Grounding Language Learning in
  Real-World Human-Machine Interaction}.
\newblock In \emph{NAACL-HLT}. Association for Computational Linguistics.

\bibitem[{Bianchi et~al.(2020{\natexlab{a}})Bianchi, Tagliabue, Yu, Bigon, and
  Greco}]{BianchiSIGIReCom2020}
Federico Bianchi, Jacopo Tagliabue, Bingqing Yu, Luca Bigon, and Ciro Greco.
  2020{\natexlab{a}}.
\newblock \href {https://arxiv.org/abs/2007.14906} {Fantastic embeddings and
  how to align them: Zero-shot inference in a multi-shop scenario}.
\newblock In \emph{Proceedings of the SIGIR 2020 eCom workshop}.

\bibitem[{Bianchi et~al.(2020{\natexlab{b}})Bianchi, Yu, and
  Tagliabue}]{bianchi2020bert}
Federico Bianchi, Bingqing Yu, and Jacopo Tagliabue. 2020{\natexlab{b}}.
\newblock Bert goes shopping: Comparing distributional models for product
  representations.
\newblock \emph{arXiv preprint arXiv:2012.09807}.

\bibitem[{Bisk et~al.(2020)Bisk, Holtzman, Thomason, Andreas, Bengio, Chai,
  Lapata, Lazaridou, May, Nisnevich, Pinto, and Turian}]{Bisk2020ExperienceGL}
Yonatan Bisk, Ari Holtzman, Jesse Thomason, Jacob Andreas, Yoshua Bengio, Joyce
  Chai, Mirella Lapata, Angeliki Lazaridou, Jonathan May, Aleksandr Nisnevich,
  Nicolas Pinto, and Joseph Turian. 2020.
\newblock \href {https://doi.org/10.18653/v1/2020.emnlp-main.703} {Experience
  grounds language}.
\newblock In \emph{Proceedings of the 2020 Conference on Empirical Methods in
  Natural Language Processing (EMNLP)}, pages 8718--8735, Online. Association
  for Computational Linguistics.

\bibitem[{Bojanowski et~al.(2017)Bojanowski, Grave, Joulin, and
  Mikolov}]{bojanowski2016enriching}
Piotr Bojanowski, Edouard Grave, Armand Joulin, and Tomas Mikolov. 2017.
\newblock \href {https://doi.org/10.1162/tacl_a_00051} {Enriching word vectors
  with subword information}.
\newblock \emph{Transactions of the Association for Computational Linguistics},
  5:135--146.

\bibitem[{Catalano et~al.(2009)Catalano, Leise, and Pfaff}]{gini}
Michael Catalano, Tanya Leise, and Thomas Pfaff. 2009.
\newblock \href {https://doi.org/10.5038/1936-4660.2.2.4} {Measuring resource
  inequality: The gini coefficient}.
\newblock \emph{Numeracy}, 2.

\bibitem[{Cramer-Flood(2020)}]{emarketer2020}
Ethan Cramer-Flood. 2020.
\newblock \href {https://www.emarketer.com/content/global-ecommerce-2020}
  {\emph{Global Ecommerce 2020. Ecommerce Decelerates amid Global Retail
  Contraction but Remains a Bright Spot.}}

\bibitem[{Craswell et~al.(2020)Craswell, Mitra, Yilmaz, Campos, and
  Voorhees}]{Craswell2020OverviewOT}
Nick Craswell, Bhaskar Mitra, E.~Yilmaz, Daniel~Fernando Campos, and
  E.~Voorhees. 2020.
\newblock Overview of the trec 2019 deep learning track.
\newblock \emph{ArXiv}, abs/2003.07820.

\bibitem[{Devlin et~al.(2019)Devlin, Chang, Lee, and Toutanova}]{bert}
Jacob Devlin, Ming-Wei Chang, Kenton Lee, and Kristina Toutanova. 2019.
\newblock \href {https://doi.org/10.18653/v1/N19-1423} {{BERT}: Pre-training of
  deep bidirectional transformers for language understanding}.
\newblock In \emph{Proceedings of the 2019 Conference of the North {A}merican
  Chapter of the Association for Computational Linguistics: Human Language
  Technologies, Volume 1 (Long and Short Papers)}, pages 4171--4186,
  Minneapolis, Minnesota. Association for Computational Linguistics.

\bibitem[{{Econsultancy}(2020)}]{econsultancy2020}
{Econsultancy}. 2020.
\newblock \href
  {https://econsultancy.com/site-search-retailers-still-have-a-lot-to-learn/}
  {\emph{Site search: retailers still have a lot to learn.}}

\bibitem[{Egg(2019)}]{Grubhub}
Alex Egg. 2019.
\newblock \href
  {https://bytes.grubhub.com/search-query-embeddings-using-query2vec-f5931df27d79?gi=c162382d0192}
  {\emph{Query2vec: Search query expansion with query embeddings}}.

\bibitem[{Grbovic et~al.(2016)Grbovic, Djuric, Radosavljevic, Silvestri,
  Baeza-Yates, Feng, Ordentlich, Yang, and Owens}]{Grbovic2016ScalableSM}
Mihajlo Grbovic, Nemanja Djuric, Vladan Radosavljevic, Fabrizio Silvestri,
  Ricardo Baeza-Yates, Andrew Feng, Erik Ordentlich, Lee Yang, and Gavin Owens.
  2016.
\newblock \href {https://doi.org/10.1145/2911451.2911538} {Scalable semantic
  matching of queries to ads in sponsored search advertising}.
\newblock In \emph{Proceedings of the 39th International ACM SIGIR Conference
  on Research and Development in Information Retrieval}, SIGIR '16, page
  375–384, New York, NY, USA. Association for Computing Machinery.

\bibitem[{Grbovic et~al.(2015)Grbovic, Radosavljevic, Djuric, Bhamidipati,
  Savla, Bhagwan, and Sharp}]{Grbovic15}
Mihajlo Grbovic, Vladan Radosavljevic, Nemanja Djuric, Narayan Bhamidipati,
  Jaikit Savla, Varun Bhagwan, and Doug Sharp. 2015.
\newblock \href {https://doi.org/http://dx.doi.org/10.1145/2783258.2788627}
  {E-commerce in your inbox: Product recommendations at scale}.
\newblock In \emph{Proceedings of KDD '15}.

\bibitem[{Karpukhin et~al.(2020)Karpukhin, Oguz, Min, Lewis, Wu, Edunov, Chen,
  and Yih}]{Karpukhin2020DensePR}
Vladimir Karpukhin, Barlas Oguz, Sewon Min, Patrick Lewis, Ledell Wu, Sergey
  Edunov, Danqi Chen, and Wen-tau Yih. 2020.
\newblock \href {https://doi.org/10.18653/v1/2020.emnlp-main.550} {Dense
  passage retrieval for open-domain question answering}.
\newblock In \emph{Proceedings of the 2020 Conference on Empirical Methods in
  Natural Language Processing (EMNLP)}, pages 6769--6781, Online. Association
  for Computational Linguistics.

\bibitem[{Landauer and Dumais(1997)}]{Landauer1997AST}
Thomas~K. Landauer and Susan~T. Dumais. 1997.
\newblock A solution to plato's problem: The latent semantic analysis theory of
  acquisition, induction, and representation of knowledge.

\bibitem[{Lewis(1969)}]{lewis1969convention}
David Lewis. 1969.
\newblock Convention.
\newblock \emph{Mass.: Harvard UP}.

\bibitem[{McHugh(2012)}]{articleKappa}
Mary McHugh. 2012.
\newblock \href {https://doi.org/10.11613/BM.2012.031} {Interrater reliability:
  The kappa statistic}.
\newblock \emph{Biochemia medica : časopis Hrvatskoga društva medicinskih
  biokemičara / HDMB}, 22:276--82.

\bibitem[{Mikolov et~al.(2013{\natexlab{a}})Mikolov, Chen, Corrado, and
  Dean}]{Mikolov2013EfficientEO}
Tomas Mikolov, Kai Chen, Gregory~S. Corrado, and Jeffrey Dean.
  2013{\natexlab{a}}.
\newblock Efficient estimation of word representations in vector space.
\newblock \emph{CoRR}, abs/1301.3781.

\bibitem[{Mikolov et~al.(2013{\natexlab{b}})Mikolov, Sutskever, Chen, Corrado,
  and Dean}]{w2vec}
Tomas Mikolov, Ilya Sutskever, Kai Chen, Greg Corrado, and Jeffrey Dean.
  2013{\natexlab{b}}.
\newblock Distributed representations of words and phrases and their
  compositionality.
\newblock In \emph{Proceedings of the 26th International Conference on Neural
  Information Processing Systems - Volume 2}, NIPS’13, page 3111–3119, Red
  Hook, NY, USA. Curran Associates Inc.

\bibitem[{Montague(1974)}]{montague1974c}
Richard Montague. 1974.
\newblock English as a formal language.
\newblock In Richmond~H. Thomason, editor, \emph{Formal Philosophy: Selected
  Papers of Richard Montague}, pages 188--222. Yale University Press, New
  Haven, London.

\bibitem[{Mu et~al.(2018)Mu, Yang, and Yan}]{DBLP:journals/corr/abs-1804-00306}
Cun Mu, Guang Yang, and Zheng Yan. 2018.
\newblock \href {http://arxiv.org/abs/1804.00306} {Revisiting skip-gram
  negative sampling model with regularization}.
\newblock \emph{CoRR}, abs/1804.00306.

\bibitem[{Nogueira et~al.(2020)Nogueira, Jiang, and
  Lin}]{Nogueira2020DocumentRW}
Rodrigo Nogueira, Zhiying Jiang, and Jimmy Lin. 2020.
\newblock Document ranking with a pretrained sequence-to-sequence model.
\newblock In \emph{EMNLP}.

\bibitem[{Nozza et~al.(2020)Nozza, Bianchi, and Hovy}]{nozza2020mask}
Debora Nozza, Federico Bianchi, and Dirk Hovy. 2020.
\newblock {What the [MASK]? making sense of language-specific BERT models}.
\newblock \emph{arXiv preprint arXiv:2003.02912}.

\bibitem[{Peters et~al.(2018)Peters, Neumann, Iyyer, Gardner, Clark, Lee, and
  Zettlemoyer}]{elmo}
Matthew Peters, Mark Neumann, Mohit Iyyer, Matt Gardner, Christopher Clark,
  Kenton Lee, and Luke Zettlemoyer. 2018.
\newblock \href {https://doi.org/10.18653/v1/N18-1202} {Deep contextualized
  word representations}.
\newblock In \emph{Proceedings of the 2018 Conference of the North {A}merican
  Chapter of the Association for Computational Linguistics: Human Language
  Technologies, Volume 1 (Long Papers)}, pages 2227--2237, New Orleans,
  Louisiana. Association for Computational Linguistics.

\bibitem[{Rogers et~al.(2020)Rogers, Kovaleva, and
  Rumshisky}]{rogers2020bertology}
Anna Rogers, Olga Kovaleva, and Anna Rumshisky. 2020.
\newblock \href {https://doi.org/10.1162/tacl_a_00349} {A primer in
  {BERT}ology: What we know about how {BERT} works}.
\newblock \emph{Transactions of the Association for Computational Linguistics},
  8:842--866.

\bibitem[{Strubell et~al.(2019)Strubell, Ganesh, and
  McCallum}]{strubell2019energy}
Emma Strubell, Ananya Ganesh, and Andrew McCallum. 2019.
\newblock Energy and policy considerations for deep learning in nlp.
\newblock In \emph{Proceedings of the 57th Annual Meeting of the Association
  for Computational Linguistics}, pages 3645--3650.

\bibitem[{Tagliabue and Yu(2020)}]{Tagliabue2020ShoppingIT}
Jacopo Tagliabue and Bingqing Yu. 2020.
\newblock Shopping in the multiverse: A counterfactual approach to in-session
  attribution.
\newblock In \emph{eCOM@ SIGIR}.

\bibitem[{Tagliabue et~al.(2020{\natexlab{a}})Tagliabue, Yu, and
  Beaulieu}]{CoveoECNLP22}
Jacopo Tagliabue, Bingqing Yu, and Marie Beaulieu. 2020{\natexlab{a}}.
\newblock \href {https://doi.org/10.18653/v1/2020.ecnlp-1.2} {How to grow a
  (product) tree: Personalized category suggestions for e{C}ommerce
  type-ahead}.
\newblock In \emph{Proceedings of The 3rd Workshop on e-Commerce and NLP},
  pages 7--18, Seattle, WA, USA. Association for Computational Linguistics.

\bibitem[{Tagliabue et~al.(2020{\natexlab{b}})Tagliabue, Yu, and
  Bianchi}]{10.1145/3383313.3411477}
Jacopo Tagliabue, Bingqing Yu, and Federico Bianchi. 2020{\natexlab{b}}.
\newblock \href {https://doi.org/10.1145/3383313.3411477} {The embeddings that
  came in from the cold: Improving vectors for new and rare products with
  content-based inference}.
\newblock In \emph{Fourteenth ACM Conference on Recommender Systems}, RecSys
  '20, page 577–578, New York, NY, USA. Association for Computing Machinery.

\bibitem[{Techcrunch()}]{CoveoRound}
Techcrunch.
\newblock \href
  {https://techcrunch.com/2019/11/06/coveo-raises-227m-at-1b-valuation-for-ai-based-enterprise-search-and-personalization/}
  {coveo-raises-227m-at-1b-valuation}.

\bibitem[{{Techcrunch}(2019{\natexlab{a}})}]{AlgoliaRound}
{Techcrunch}. 2019{\natexlab{a}}.
\newblock \href
  {https://techcrunch.com/2019/10/15/algolia-finds-110m-from-accel-and-salesforce-for-its-search-as-a-service-used-by-slack-twitch-and-8k-others/}
  {Algolia finds \$110m from accel and salesforce}.

\bibitem[{{Techcrunch}(2019{\natexlab{b}})}]{LWRound}
{Techcrunch}. 2019{\natexlab{b}}.
\newblock \href
  {https://techcrunch.com/2019/08/12/lucidworks-raises-100m-to-expand-in-ai-powered-search-as-a-service-for-organizations/}
  {Lucidworks raises \$100m to expand in ai finds}.

\bibitem[{Tsagkias et~al.(2020)Tsagkias, King, Kallumadi, Murdock, and
  de~Rijke}]{Tsagkias2020ChallengesAR}
Manos Tsagkias, Tracy~Holloway King, Surya Kallumadi, Vanessa Murdock, and
  Maarten de~Rijke. 2020.
\newblock Challenges and research opportunities in ecommerce search and
  recommendations.
\newblock In \emph{SIGIR Forum}, volume~54.

\bibitem[{Vasile et~al.(2016)Vasile, Smirnova, and
  Conneau}]{Vasile2016MetaProd2VecPE}
Flavian Vasile, Elena Smirnova, and Alexis Conneau. 2016.
\newblock Meta-prod2vec: Product embeddings using side-information for
  recommendation.
\newblock \emph{Proceedings of the 10th ACM Conference on Recommender Systems}.

\bibitem[{Yu and Tagliabue(2020)}]{Yu2020BlendingSA}
Bingqing Yu and Jacopo Tagliabue. 2020.
\newblock Blending search and discovery: Tag-based query refinement with
  contextual reinforcement learning.
\newblock In \emph{EComNLP}.

\bibitem[{Yu et~al.(2020)Yu, Tagliabue, Greco, and Bianchi}]{Yu2020AnII}
Bingqing Yu, Jacopo Tagliabue, Ciro Greco, and Federico Bianchi. 2020.
\newblock An image is worth a thousand features: Scalable product
  representations for in-session type-ahead personalization.
\newblock \emph{Companion Proceedings of the Web Conference 2020}.

\end{thebibliography}
\bibliographystyle{acl_natbib}

\end{document}